\documentclass[12pt]{article} 
\def \no{\noindent}

\begin{document}

\LARGE

\centerline{\bf A new line element derived from the variable}
\centerline{\bf rest mass in gravitational field
\footnote{~The material in this paper was partially presented in the IARD
 conferences, June~2002 and June~2006}}

\normalsize
\bigskip
\centerline{\bf N. Ben-Amots}
\medskip

\centerline{P.O.Box 3193, Haifa 31032, Israel}

\centerline{e-mails: benamots@alumni.technion.ac.il}
\centerline{nbenamots@yahoo.com}

\setlength{\parindent}{1cm}

\begin{quotation}

{\it This paper presents a new line element based on the assumption of the 
variable rest mass in gravitational field, and explores some 
its implications. This line element is {\rm not} a {\rm vacuum} 
solution of Einstein's equations, yet it is sufficiently close to 
Schwarzschild's line element to be compatible with all of the 
experimental and observational measurements made so far to confirm 
the three Einstein's predictions.  The theory allows radiation and 
fast particles to escape from all massive bodies, even from those 
that in Einstein's general relativity framework will be black holes.  
The striking feature of this line element is the non-existence of 
{\rm black holes}.}

\medskip

        {\bf KEY WORDS:} gravitation; gravitational
potential; variable rest mass, black holes.
\end{quotation}

\section{Introduction}
In Newtonian physics the final state of the gravitational collapse
is a state of infinite mass density not separated from the outside
word by a horizon of events and its formation is accompanied by
radiating an infinite amount of heat. The general relativity offers
the possibility of stable end state of collapsing mass body called
black hole representing a mass object from which electromagnetic radiation 
cannot escape. Besides the general feeling of physicists who have a
psychological resistance against the existence of any singular state
of matter in both cases, one also faces difficulties with the
physical interpretation of the final product of collapsing stars. The
classical naked singularity represents a highly unphysical state of
matter and the existence of black hole in its general form is
conditioned with the cosmic censorship hypothesis that appears not
generally proved so far. Therefore, the endeavor to find alternatives
to the black holes seems still justified. In what follows, we
will investigate the possibility of the black hole by applying the 
principle of energy conservation to the gravitational interaction.

In this paper we consider the possibility of variation of rest mass due to
gravitational fields.

As is well known, the {\it energy} is one of the most fundamental
concepts of physics and its conservation is evident in all its
subdisciplines. This is why we apply the principle of energy
conservation to the motion of the test particle in gravitational
field by allowing that its mass in a gravitational field is
variable. With the assumption that the total energy of a test particle
in gravitational field is given by the sum of the energy equivalent
of the mass of the particle at infinity plus its potential gravitational 
energy we have derived a new line element and show that it is in agreement 
with all Einstein's observational tests of general relativity results so far.

In relativity, every potential energy~$\Delta E$ of a body, except the
gravitational potential energy, increases its mass by a quantity
$$\Delta m = \Delta E/c^2 \eqno(1)$$

Instead of using the potential energy
of gravity as in the Newtonian theory, Einstein's general relativity uses
the concept of space being distorted by mass.  When considering, as a way
of convenience, the gravitational potential energy instead of curved space,
most authors do not allow the gravitational potential energy to contribute
to the mass of a body.  The energy~$\Delta E$ 
of an electric charged particle near
another electric charge, varies depending on the distance between them.
According to Equation~$(1)$, the rest mass of the charged particle varies
in the  same measure.  This happens in the same manner for all
non-gravitational energies.  Yet when a body changes its potential energy
in a gravitational field, general relativity conserves the rest mass of the
body.  Thus general relativity gives gravity a unique status.

In the following, we do allow for {\it mass variation due to gravitational
potential energy}, and explore some consequences.  

In classical physics or general relativity one finds the force of
electrostatic attraction between two static electric charges by calculating
the derivative of the potential energy with respect to the distance between
the charges.  Similarly we calculate the gravitational force between two  
bodies as the derivative with respect to the distance of 
a Newtonian gravitational potential, in which the rest masses are variable.
The justification for the derivative with distance
is that in this quasistatic approximation, the momentum part 
contribution to~$E$ is zero.  Our results pass all the experimental tests 
that the three Einstein's predictions for general relativity also passes, at 
the present level of precision.

In the literature, there are many different theories with variable
rest mass, some of which, e.g.\ [1]-[7] do not lead to  Einstein's
field equations. See also [8]-[11] and in particular Einstein and
Infeld [12] as well as the review of Szabados [13].

The theory presented here is  different from the exponential
gravitation theory [14]-[27] in which the rest mass is variable as
well.

 \section{Variable rest mass of test particle }

\noindent
The total energy (mass energy when isolated at infinity + potential 
gravitational energy) of a static particle of mass~$m$ in a Newtonian 
gravity field caused by another large central mass~$M$ located at the 
origin of a spherical coordinate system is:
$$ E = m_{\infty}c^2-\frac{GmM}{r} \eqno(2)$$
\noindent
where~$G$ is the universal gravitational constant, and~$r$ is the distance
of~$m$ from the origin where $M$ is situated.
Equation~$(2)$ simply states that the energy of a static 
particle~$m$ at a distance~$r$ from a central mass~$M$, is the energy
of its rest mass at infinite distance, plus the negative gravitational
energy (potential energy) at distance~$r$ (where its mass~$m$ is allowed to
be variable).  For the gravitational energy we substituted the Newtonian 
gravitational potential energy but with a variable mass $m$.

We think that
an energy decrease of $GmM/r$ as an approximation to the inclusion of
gravitation in the energy of a body, is more realistic than the assumptions of general 
relativity, and closer to Einstein and Infeld's [12] and Arnowitt et~al.'s [1] descriptions.

\medskip
Given the assumption that the mass {\it does} depend on gravitational
potential energy as well, we get for the mass of the particle:
$$ m = \frac{E}{c^2} = m_{\infty}-\frac{GmM}{c^2r}  \eqno(3)$$

In~$(3)$ at infinite distance~$r$ the mass~$m$ becomes~$m_{\infty}$ as
it should.

From~$(3)$ we find the mass~$m$ and the energy~$E$ of the particle:
$$ m(r) = \frac{m_{\infty}}{1+GM/c^2r}  \eqno(4)$$

Later when discussing Equation~$(11)$ we explain what it means for
$m(r)$ to vanish when~$r$ approaches zero.

From~$(4)$ we get:
$$ E = mc^2 = \frac{m_{\infty}c^2}{1+GM/c^2r}  \eqno(5)$$

The mass~$m$ depends on~$r$, while temporarily we take the mass~$M$ as
independent of~$r$.  Later we will also give some consideration on variable
mass~$M$.

We consider the {\it quasistatic} case of gravitation interaction
between two bodies in which the approximate gravitational force is:
$$ \vec{F} = -\frac{dE}{dr} = -\frac{Gm_{\infty}M}{(r+
GM/c^2)^2}\left[\frac{\vec{r}}{r}\right]  \eqno(6)$$

If ~$r \longrightarrow \infty$ we get~$E=m_{\infty}c^2$. When
~$r>>GM/c^2$, then ~$F$ approaches $Gm_{\infty}M/r^2$ in accord
with the Newtonian gravitation theory.

By~$(4)$,~$(5)$, the energy or mass of a static particle at a distance~$r$ 
to the central mass~$M$, approaches zero when~$r$ approaches zero.  

Schwarzschild's line element of general relativity is
(for example Adler et~al.\ [28] p.\ 194, Equation~$(6.53)$):
$$ ds^2 = c^2dt^2\left(1-\frac{2GM}{c^2r}\right)+\frac{dr^2}{1-2GM/c^2r}
-r^2d\theta^2-r^2\sin^2{\theta}d\phi^2  \eqno(7)$$

Although our assumption for getting the results~$(2)-(6)$ is not based
on any special geometry (except the Euclidean,) if, {\it only as a means of
comparison} with general relativity, we were to describe our equations by 
means of a line element in a curved space as general relativity does, the 
corresponding line element would be:
$$ ds^2 = \frac{1-GM/rc^2}{1+GM/rc^2}c^2dt^2-
\frac{1+GM/rc^2}{1-GM/rc^2}dr^2-
r^2d\theta^2-r^2\sin^2{\theta}d\phi^2  \eqno(8)$$

The coefficient~$\frac{1-GM/rc^2}{1+GM/rc^2}$ of~$c^2dt^2$~$(8)$
differs from the coefficient~${1-2GM/rc^2}$ in~$(7)$ only in orders higher
than the first order of~$GM/rc^2$.

We derived~$(8)$ using the procedure followed by Weinberg [29] 
who, while deriving Schwarszchild's line element~$(7)$ showed
that $g_{00}=1+2\Phi$,
where he takes~$\Phi$ as the Newtonian gravity potential~$-GM/rc^2$.

In our case (variable rest mass) we derived~$\Phi$ by
subtracting~$m_{\infty}c^2$ from~$(5)$, then dividing the result
by~$m_{\infty}c^2$ and we received~$\Phi$ equal to~$1/(1+GM/rc^2)-1$.
Calculating~$1+2\Phi$ we get the
coefficient~$(1-GM/rc^2)/(1+GM/rc^2)$, and our "line element"~$(8)$.

We believe, however, that Weinberg's procedure is an approximation, and
that a very close but better pseudo line element
is achieved if the factor~2 is introduced 
not as in $2\phi$ as in Weinberg's [29] procedure, but later, thus slightly
modify~$(8)$ to the form~$(9)$ below,
and the correct "line element" is:
$$ ds^2 = \frac{c^2dt^2}{1+2GM/rc^2}-\left(1+\frac{2GM}{rc^2}\right)dr^2-
r^2d\theta^2-r^2\sin^2{\theta}d\phi^2  \eqno(9)$$

\section{Two bodies both with variable rest mass}

\noindent
While the mass of the particle~$m$ varies with~$r$ in the gravity field of
the central mass~$M$ as expressed in~$(3)$,~$M$ also should vary in the
gravity field of~$m$:
$$ M = M(r) = M_{\infty}-\frac{GmM}{c^2r}  \eqno(10)$$

Solving the pair of the two equations~$(3)$ and~$(10)$ for~$m$ and~$M$
and summing up the solutions we get:
$$ M + m = -\frac{c^2r}{G}+\sqrt{\left[\frac{c^2r}{G}\right]^2
+2\frac{c^2r}{G}(M_{\infty}+m_{\infty})+(M_{\infty}-m_{\infty})^2} \eqno(11)$$

$M+m$ decreases when the distance~$r$ between the two masses is
decreased quasi-statically.  Particularly, when the distance~$r$ decreases
to zero the {\it sum}~$(11)$ of the masses~$M$ and~$m$ {\it decreases}
to $M_{\infty}-m_{\infty}$, so that mass equal to
2$m_{\infty}$ is "annihilated," being transformed into energy.

For two {\it equal} masses~$m$ we get the {\it approximation}:
$$ m = \frac {1}{2}\left[-\frac{c^2r}{G}+
\sqrt{\left(\frac{c^2r}{G}\right)\left(\frac{c^2r}{G}+
4m_{\infty}\right)}\:\right] \eqno(12)$$

The masses~$m$ approach zero when~$r\longrightarrow 0$, that is, the
two masses annihilate being transformed into energy.

From~$(12)$ we get:
$$ m \frac{c^2r}{2G}\left[-1+\sqrt{1+\frac{4Gm_{\infty}}{c^2r}}\:\right]=
\frac{c^2r}{2G}\left[-1+1+\frac{2Gm_{\infty}}{c^2r}+...\right]  \eqno(13)$$
\indent For~$r$ approaching infinity~$(13)$ gives
$m(r={\infty}) = m_{\infty}$ as it should.

\section{Escape velocity}

\noindent
To calculate the approximate escape velocity, we equalize the energy of a 
static particle at infinity with its approximate energy at radius~$r$ where 
it has an escape velocity~$v$:
$$E=\frac{m_{\infty}c^2}{\left(1+GM/c^2r\right)\sqrt{1-
v^2/c^2}}=m_{\infty}c^2 \eqno(14)$$

Solving~$(14)$ for v we get the escape velocity as:
$$v=c\sqrt{1-\frac{1}{(1+GM/c^2r)^2}} \eqno(15)$$

Solving~$(14)$ for~$r$ we get the smallest radius from which a particle
whose radial velocity outwards is equal to~$v$ can escape:
$$r=\frac{MG}{1/\sqrt{1-v^2/c^2}-1} \eqno(16)$$

For radial velocity outwards approaching the velocity of light~$c$,
the radius~$(16)$ approaches zero.

For large radius~$r$ it is convenient to write~$(15)$ as
$$v=\sqrt{\frac{GM}{r}}\:\frac{\sqrt{2+GM/c^2r}}{1+GM/c^2r}
 \eqno(17)$$

\noindent
and we see that for very large~$r$ we approach the Newtonian escape velocity:
$$v\approx\sqrt{\frac{2GM}{r}} \eqno(18)$$

For small radius~$r$ it is convenient to write~$(15)$ as
$$v=c\frac{\sqrt{1+2c^2r/GM}}{1+c^2r/GM} \eqno(19)$$

	From $(17)$ and $(19)$ we see that the escape velocity is always smaller 
than the velocity of light~$c$.  Thus our method, or interpretation, leads to 
continuous space without any forbidden regions, and in 
particular, without a black hole,
that is, light can escape from any radius (see~$(16)$).  A particle with
a radial outward velocity~$v$ between the values:
$$c>v>c\sqrt{1-\frac{1}{(1+GM/c^2r)^2}} \eqno(20)$$
\noindent
will escape from any distance~$r$ from~$M$, that is from~$r$ larger than zero (compare to~$(16)$).

	Schwarzschild solution of
    general relativity assumes a null Ricci tensor, which represents 
massless vacuum.  Yet a black hole is {\it not} a massless vacuum.  Thus
general relativity is not applicable in this sense; still people use it for
black holes.

	It is believed now that each galaxy has a supermassive body in its
center, including the milky way and Andromeda Galaxy
\no (Gebhardt [30],
Sch\"{o}del et~al.\ [31],
Ghez et~al.\ [32],
Genzel et~al.\ [33],
and Narayan [34]).

	It is interesting to know whether these celestial bodies of a mass
of~10$^6$~-~10$^9 M_{Sun}$, which general relativity theories show that
are black holes candidates, are indeed completely black (and do not allow
light or particles to escape from radii smaller than Schwarzschild's
radius~$r_s=2GM/c^2$), or, by contrast, are almost black and allow some
amount of radiation and particles to escape from radii less than
Schwarzschild's radius, (as the theory presented here allows).
Observations of infrared radiation from the supermassive body at the
center of our galaxy were already made ([32], [33] and Ghez~et~al.\ [35]).
Ghez~et~al.\ [35] suggest that this observed infrared radiation may be related to 
X-ray radiation around black hole.

Where particles with velocity less than~$c$ can
escape, (see~$(16)$) radiation escapes but may be
gravitationally redshifted ($(32)$ below).  Our
present theory here allows for radiation to escape from all
supermassive bodies.  Accretion around black holes
of general relativity produces mainly long-term X-ray
radiation, (which is different from the so-called Hawking radiation 
of black holes that is
negligible until a last, short final explosion).  In
our theory accretion around a supermassive body also
produces mainly X-ray radiation, but radiation that 
originates in the supermasive body itself 
can also exist and
escape and is
gravitationally redshifted $(32)$:  For example, for
radiation originating from a supermassive body
of~10$^9M_{Sun}$ that is transparent from a radius
greater than 50000km, X-rays escape as red/infra-red
radiation, and radiation with wavelength of visible
light escapes as sub-millimeter wavelength radiation.
If supermassive bodies appear to shine
red/infra-red radiation, as was found for the
massive body in the center of the Milky Way [32-35]
then this radiation may originate in the supermassive bodies.
Genzel~et~al.\ [33] "report high-resolution infrared observations
of Sgr A* that reveal 'quiescent' emission..." that they interpret
as originating in accreting gas, but we think it could be from the
radiation emitted by the central body, because emission by the central 
body can also explain the observed 'quiescent' emission.
Genzel~et~al.\ [33], Ghez~et~al.\ [32], [35] and others consider 
these supermassive bodies as black holes that cannot
emit radiation at all.  General relativity does not explain
such quiescent radiation (unless by accretion of gas), while our theory 
allows for this kind of observations originated in the central body.

\section{The three basic tests of relativistic gravitation}

In order to compare the new line element with other existing line
elements we present briefly the classical Einsteinian calculation
following  Tolman's book [36] using sometime different notations,
e.g.\ we use explicitly ~$G/c^2$ in the line elements, etc.  We
assume the general line element 
$$ds^2=-e^{\Gamma}dr^2-r^2d\theta^2-r^2\sin^2\theta
d\phi^2+e^{\mu}dt^2. \eqno(21)$$   
Following Tolman [36] p.\ 207
we get  the following three equations:
$$e^{\Gamma}\left(\frac{dr}{ds}\right)^2+
r^2\left(\frac{d\phi}{ds}\right)^2-
e^{\mu}\left(\frac{dt}{ds}\right)^2+1=0 \eqno(22)$$
$$\frac{d\phi}{ds}=\frac{H}{r^2} \eqno(23)$$
$$\frac{dt}{ds}=k e^{-\mu} \eqno(24)$$
where $k$ and $H$ are as defined by Tolman~[36].

\indent Substituting Schwarzschild's solution~$(25)$ in
Equations~$(22)$~--~$(24)$ we have
$$e^{\mu}=e^{-\Gamma}=1-\frac{2GM}{c^2r}. \eqno(25)$$
\indent The motion of an orbiting planet is described by the equation
$$\frac{\left(\frac{dr}{ds}\right)^2}{1-2GM/c^2r}+
r^2\left(\frac{d\phi}{ds}\right)^2- \frac{k^2}{1-2GM/c^2r}+1=0.
\eqno(26)$$ 
\indent Multiplying Equation~$(26)$ by the denominator for
\noindent $\frac{2GM}{c^2}<<r$ one obtains:
$$\left(\frac{dr}{ds}\right)^2+
r^2\left(\frac{d\phi}{ds}\right)^2 \left(1-\frac{2GM}{c^2r}\right)
-k^2+1-\frac{2GM}{c^2r} = 0 \eqno(27)$$ 
\indent Rearranging~$(27)$ one
receives Tolman's equation~$(83.10)$ "as (one of) the relativistic 
equations for the motion of a planet":
$$\left(\frac{dr}{ds}\right)^2+
r^2\left(\frac{d\phi}{ds}\right)^2-
\frac{2GM}{c^2r}\left[1+r^2\left(\frac{d\phi}{ds}\right)^2\right]
=k^2-1. \eqno(28)$$ 
\indent
For the comparison we present the corresponding
Newtonian equation
$$\left(\frac{dr}{ds}\right)^2+
r^2\left(\frac{d\phi}{ds}\right)^2- \frac{2GM}{c^2r}={\rm constant}
\eqno(29)$$ 
that does not contain the relativistic term mentioned by Tolman [36]
$$+\frac{3GM}{c^2r^2}\quad. \eqno(30)$$  
\indent The relativistic equation $(28)$
leads to a deflection of~1.75 seconds of arc for a ray of light near
the sun. The deflection of light and the advance of the perihelion
are both results of the same equations.

As is well known the prediction of Einstein's general relativistic
redshift is
$$\frac{L+\delta L}{L}=\frac{\delta t}{\delta s}=\frac{1}{\sqrt{e^{\mu}}}
\eqno(31)$$ 
where $L$ is wavelength.

\indent For Schwarzschild's solution, we substitute~$\mu$
from~$(25)$ and obtain:
$$\frac{L+\delta L}{L}=\frac{1}{\sqrt{1-2GM/c^2r}}
\approx 1+\frac{GM}{c^2r} \eqno(32)$$ 
\noindent that is the
gravitational shift of spectral lines. Einstein's general relativity
predictions were tested by experimental observations and found to be within 
the experimental error limits.

\section{$\!\!\!\!$Relativistic effects calculated from the proposed line
element}

In this section we calculate physical values for the three crucial observational tests
from line element $(9)$ of our theory for the quasistatic case.
For the proposed line element $(9)$ it holds
$$e^{\mu}=e^{-\Gamma}=\frac{1}{1+2GM/c^2r} \eqno(33)$$
and
$$e^{-\mu}=e^{\Gamma}=1+\frac{2GM}{c^2r}. \eqno(34)$$
\indent The gravitational shift in spectral lines is obtained by
substituting~$(33)$ in~$(31)$:
$$\frac{L+\delta L}{L}=\sqrt{1+\frac{2GM}{c^2r}}\approx 1+\frac{GM}{c^2r}
\eqno(35)$$ 
that also fits well the observational value.
Inserting~$(33)$~--~$(34)$ into~$(22)$~--~$(24)$ we obtain
$$\left(\frac{dr}{ds}\right)^2\left(1+\frac{2GM}{c^2r}\right)+
r^2\left(\frac{d\phi}{ds}\right)^2-
\frac{1}{1+2GM/c^2r}\left(\frac{dt}{ds}\right)^2+1=0 \eqno(36)$$ 
and
$$\frac{dt}{ds}=k\left(1+\frac{2GM}{c^2r}\right) \eqno(37)$$
\indent Inserting~$(37)$ into~$(36)$ one gets
$$\left(\frac{dr}{ds}\right)^2\left(1+\frac{2GM}{c^2r}\right)+
r^2\left(\frac{d\phi}{ds}\right)^2-
k^2\left(1+\frac{2GM}{c^2r}\right)+1=0 \eqno(38)$$ 
\indent Multiplying equation~$(38)$ with~$1-2GM/c^2r$ we have
$$\left(\frac{dr}{ds}\right)^2\!+
r^2\!\left(\!\frac{d\phi}{ds}\!\right)^2\!-
\frac{2GM}{c^2r}\!\left[1\!+r^2\!\left(\!\frac{d\phi}{ds}\!\right)^2\right]\!+\!
\left(\frac{2GM}{c^2}\right)^2\!\left[k^2\!-\!\left(\!\frac{dr}{ds}\!\right)^2\right]=
k^2-1\eqno(39)$$ 
\indent Neglecting the higher order terms
of~$\left(2GM/c^2r\right)^2$ we obtain
$$\left(\frac{dr}{ds}\right)^2+
r^2\left(\frac{d\phi}{ds}\right)^2-
\frac{2GM}{c^2r}\left[1+r^2\left(\frac{d\phi}{ds}\right)^2\right]
=k^2-1 $$ that is exactly the familiar equation~$(28)$, that leads
to corrected value of the deflection of light.

Instead of multiplying Equation~$(38)$ with~$1-2GM/c^2r$, it is
preferable to multiply~$(38)$ with
$$\frac{1}{1\!+\!2GM/c^2r}\!=\!
\frac{1\!+\!2GM/c^2r\!-\!2GM/c^2r}{1+2GM/c^2r}\!=\!
1\!-\!\frac{2GM/c^2r}{1\!+\!2GM/c^2r} \eqno(40)$$ that gives after
rearranging:
$$\left(\frac{dr}{ds}\right)^2+
r^2\left(\frac{d\phi}{ds}\right)^2-
\frac{1+r^2\left(\frac{d\phi}{ds}\right)^2}{1+c^2r/2GM}
=k^2-1 \eqno(41)$$

This is an approximation that includes also higher order terms than
equations~$(28)$ or~$(39)$. It is valid in any gravitational field.
This should lead to predictions closer to the observational value for 
the deflection of light in a gravitational field as well as for the 
advance of the perihelia of the planets.  Similar results were 
obtained by Coleman [20], and by Majern\'{\i}k [37]  with the 
exponential gravitation metric.

\section{Comparing line elements by Schiff's procedure}

When developed in series,~$(9)$ agrees with Schwarzschild's line
element~$(7)$ to the first order in~$2GM/(c^2r)$, yet they differ in higher
orders.

Schiff [38],[39] analyzed Schwarzschild's solution~$(7)$, and
concluded that "higher terms in the series
$$ds^2=\left(1+\alpha\frac{m}{r}+\beta\frac{m^2}{r^2}+...\right)c^2dt^2
-\left(1+\gamma\frac{m}{r}+\delta\frac{m^2}{r^2}+...\right)dr^2
-r^2(d\theta^2+\sin^2\theta d\phi^2) \eqno(42)$$
\noindent
have not been subjected to experimental test."  To the best of our 
knowledge, no reliable experimental data were found to contradict Schiff's
remark until this very day.

     We compare Schiff's general line element~$(42)$ with other line
elements:

     We substitute~$2GM/c^2r$ instead~$m/r$ in~$(42)$.

     A line element far from a mass should be asymptotically equal to the
flat-space line element.  For any line element asymptotically equal to the
flat-space line element one condition is that the coefficient of~$c^2dt^2$
multiplied by the coefficient of~$dr^2$ is~-1 (See Adler~et~al.\ [28]
p.\ 192).  Using this we may reduce Schiff's line element $(42)$ as
function of $2GM/c^2r$ to:
$$ds^2\!=f\left(\frac{2GM}{c^2r}\right)c^2dt^2
-\frac{dr^2}{f(2GM/c^2r)}
\!-r^2(d\theta^2\!+\sin^2\!\theta d\phi^2) \eqno(43)$$

\noindent
where for a line element asymptotically equal to the flat-space line
element
$$f\left(\frac{2GM}{c^2r}\right)=1-1\left(\frac{2GM}{c^2r}\right)+
p\left(\frac{2GM}{c^2r}\right)^2+... \eqno(44)$$

\noindent
and~$p$ and higher coefficients have not been subjected to experimental
test.

For flat space line element:
$$f\left(\frac{2GM}{c^2r}\right)=1 \eqno(45)$$

For Schwarzschild's solution~$(7)$:
$$f\left(\frac{2GM}{c^2r}\right)=1-1\:\left(\frac{2GM}{c^2r}\right) \eqno(46)$$

For our "Line Element"~$(9)$:
$$f\left(\frac{2GM}{c^2r}\right)=\frac{1}{1+2GM/c^2r}=1
-1\left(\frac{2GM}{c^2r}\right)
+1\left(\frac{2GM}{c^2r}\right)^2+... \eqno(47)$$

Equations~$(46)$ and~$(47)$ show that both the Schwarzschild line
element and our "line 
element"~$(9)$ are identical to the first order of~$2GM/c^2r$.

Actual measurements have been compatible until now with
both~$(7)$, and~$(9)$ and in fact with any solution that does not differ
from Schwarzschild's solution~$(7)$ in the {\it first} order of~$2GM/c^2r$.

The two theories differ from each other in higher terms in Schiff's
expansion~$(42)$.  Even today these higher terms could not have been
measured experimentally because they are much smaller than the accuracy of
measurement of the lower order term.

We remark that confusion can arise from the fact that~$\beta$ in~$(42)$
can be mistaken for~$\beta$ in PPN (Parametrized Post Newtonian) notation
that is used for similar purpose, but defined differently.  In PPN
notation,~$\beta$=1 for general relativity, while the observational
measurements of the advance of the perihelion of Mercury set an upper limit
of~3$\times$10$^{-3}$ to~$\beta_{PPN}$-1.  See Will [40] Table 4.
For~$r=r_{Mercury}$ given as the radius of Mercury's orbit around the sun,
this upper limit is equivalent to
$$\beta_{Schiff}<\frac{
3\times10^{-3}c^2r_{Mercury}}{2GM_{Sun}}\approx2\times10^4 \eqno(48)$$

We consider methods to measure the advance of perihelion more accurately
than by Mercury's orbit measurements.
This may yield
observational values for the coefficient $p$ of $\left(\frac{2GM}{c^2r}\right)^2$
in~$(44)$, or an upper
limit for it.  If it is found that~$p$ in~$(44)$ is zero, it will confirm
general relativity and disprove the theory presented here (that
predicts~$p=1$, which is equivalent to
$\beta_{PPN}-1\approx-1.5\times10^{-7}$ well below the upper limit
$\beta_{PPN}-1=3\times10^{-3}$ measured from the advance of
perihelion of Mercury.  See Equation~$(48)$).
(Exponential gravitation theory [14] - [27] predicts $p=\frac{1}{2}$, or
$\beta_{PPN}-1\approx-0.75\times10^{-7}$).
The necessary accuracy is 20000 (=3$\times$10$^{-3}$/1.5$\times$10$^{-7}$) times 
greater than the accuracy achieved by Mercury measurements.

Non-Mercury planets in the solar system have orbits of smaller
eccentricity and longer periods than Mercury.  Their advance of perihelion
cannot be measured sufficiently accurate as that of Mercury, let alone to
find higher coefficients discussed in this paper.

The orbits of Earth-orbit crossing asteroids (abbreviated NEO and NEA)
are accurately measured to find any danger of collision with Earth.  There
exist databases of precision data of their orbits 
[41]-[47].  Some of these orbits are more
eccentric than Mercury, and some of them has periods of only a year or less.
Icarus's rate of relativistic advance of perihelion is less than a quarter of 
that of Mercury (Gilvarry [48],[49]), so when checking the databases of asteroids 
[41]-[47] one needs to find an asteroid whose relativistic advance of perihelion 
is 100000 times greater than that of Icarus, in hope that its measurements will
yield results 20000 times more accurate than those of Mercury's measurements.

La Paz [50] considered Earth's satellites for accurate measurements of their
relativistic advance of perihelion.  His conclusions point that a satellite orbiting with
radius of 7200 km from Earth's center may have a larger rate of advance of perihelion, 
but not 20000 times that of Mercury.  Actually the
measurements of Mercury's orbit yielded more accurate data for calculating
$\beta_{PPN}$ then the data available from Earth's artificial satellites.

Gilvarry [51] suggested what he called artificial planets orbiting the Sun (that is,
Sun artificial satellites), for accurately measuring their advance of perihelion.

The approximate relativistic advance of perihelion from one orbit to the next is:
$$\Delta\phi=\frac{6\pi G}{c^2}\frac{M}{r_-} \eqno(49)$$
where $r_-$ denotes the minimal radius during an elliptic orbit (perihelion), and 
$M\!=\!M_{Sun}$ for planets and for Sun satellites.
For Mercury $\Delta\phi$ results about 0.103" in one period.

Considering Gilvarry's [51] Sun satellite orbit having a small eccentricity and a radius 
of about one million km from Sun's center, $\Delta\phi$ results about 55.5 larger than 
that of Mercury.  Even if a single orbit of Sun's satellite could be measured 
accurately as the orbit of Mercury, 55 is far less than the accuracy needed for
theory comparison
of 20000 times greater than that of Mercury measurements.
Yet calculating the {\it rate} of advance of perihelion, 
that is the advance of perihelion during a certain time,
one may take advantage of the shorter
period of this Sun's satellite according to Kepler's law.
Considering this, the rate is proportional to $1/[(r_-)(r_+)^{3/2}]$, where $r_+$ denotes
the maximal radius (aphelion) during an elliptic orbit.
For nearly circular orbits we may approximate $r_-\!\approx\!r_+\!\approx\!r$, giving 
rate of relativistic perihelion proportional to
$1/r^{5/2}$.  So the rate of advance of perihelion of closely circular orbit of Sun's 
satellites one million km distant from the Sun's center will be (55.5)$^{2.5}\!\!\approx$23000
greater than that of Mercury (that is $\approx$273$^o$ per century. For radius 
$r\!\approx$0.9 million km, the advance of the perihelion is almost one rotation per century).
If the accuracy obtained indeed will be 23000 greater than that achieved so far by 
measurements of Mercury's rate of advance of perihelion, this may be sufficient to
determine which theory fits better the measurements.

Yet, measurements of the advance of perihelion are very difficult for almost 
circular orbits.  Possible satellite orbits around the Sun were calculated to 
find the optimal parameters for precise
measurements [52], [53].  The optimum for measurement purpose is
achieved for eccentricity of 0.816 [52].  The minimum perihelion radius $r_-$ must be larger 
than the radius of the Sun plus at least few hundred thousand km.  This means 
that the large aphelion radius $r_+$ of the ellipse should be about 10 times larger than the small
perihelion radius $r_-$.  Thus, the orbiting period of the advance of perihelion of an optimal 
elliptic orbit is about 30 times longer than that of almost circular orbit.  
Considering all this, the expected precision 
by which the advance of perihelion can be measured
is about 800 times better than the nowadays precision of 
measurement of Mercury's advance of perihelion.
This is still about 25 times worse than required 
for checking candidate theories.  Also, difficulties will exist in withstanding the 
high temperature, solar winds and strong magnetic field near the Sun, and 
measurement difficulties because of solar winds and oblateness of the Sun, etc. 
See also [53], [54].  Hopefully the discrepancy 
between required precision and present available precision
and the mentioned difficulties will be overcome in future, and 
an adequate artificial satellite with optimal orbit will
enable determination of the theory that better fits the observations.

The actual calculation should use more accurate analytical solution of the relativistic
advance of perihelion, than that obtained by Einstein.  Saca [55] indeed presented 
a more accurate formula than that of Einstein, for the general relativistic advance of
perihelion, and calculated and compared the results for Mercury and Icarus [56].
According to Saca's [56] corrected formula the advance of perihelion of Mercury is 42.981244" 
per century, compared to 42.981236" per century according to Eintstein's formula~$(49)$.

So, when such Sun's artificial satellite will orbit the sun and its rate of
relativistic advance of perihelion be measured, 
it is
hopefully plausible that the coefficient $p$ in $(44)$ 
can be calculated with sufficient accuracy, after
the observational
measurements should be checked whether they fit Saca's 
general-relativistic formula [55] ($p$=0), or the formula calculated
from the theory in this paper ($p$=1), or the exponential gravitation theory [14] - [27] ($p$=1/2),
thus determining whether the value of $p$ in $(44)$  is 0 or 1 or 1/2, respectively.

\section{Remarks}

\begin{enumerate}
\item [i)]  The Schwarzschild line element represents the vacuum
solution of Einstein's equation. Since the gravitational field
surrounding the central mass body includes field energy that itself is a
source of gravitational field the Schwarzschild vacuum solution
represents only a very good approximation to a more realistic one.
\item [ii)]  The proposed line element does not form a horizon of events,
which implies that the  black hole does not exist in the present theory.
\item [iii)]  It seems that only future astronomical measurements,
for example of near-Sun artificial satellites, can decide which of the 
presented line elements is compatible with observation.
\end{enumerate}

\section{Acknowledgements}
\noindent The author is deeply indebted to Prof.\ Vladim\'{i}r Majern\'{\i}k who
has read the paper, has done many important remarks, has given a number of 
valuable propositions, and helped edit the paper.

The author thanks Dan Igner, who helped find the appropriate wording for a clearer 
presentation.

\end{document}